\documentclass{article}

\usepackage[utf8]{inputenc}
\usepackage[left=3cm, right=3cm, top=2cm, bottom=2cm]{geometry}

\usepackage{subfiles}

\usepackage[style=numeric,backref=true,url=false]{biblatex}
\usepackage{hyperref}
\DefineBibliographyStrings{english}{
	backrefpage = {page},
	backrefpages = {pages},
}
\input{commands}

\title{Introducing Proof Tree Automata and Proof Tree Graphs}
\author{Valentin D. Richard}
\date{2022}

\begin{document}

\maketitle

\begin{abstract}
	In structural proof theory, designing and working on large calculi make it difficult to get intuitions about each rule individually and as part of a whole system. We introduce two novel tools to help working on calculi using the approach of graph theory and automata theory. The first tool is a Proof Tree Automaton (PTA): a tree automaton which language is the derivation language of a calculus. The second tool is a graphical representation of a calculus called Proof Tree Graph (PTG). In this directed hypergraph, vertices are sets of terms (e.g. sequents) and hyperarcs are rules. We explore properties of PTA and PTGs and how they relate to each other. We show that we can decompose a PTA as a partial map from a calculus to a traditional tree automaton. We formulate that statement in the theory of refinement systems. Finally, we compare our framework to proof nets and string diagrams.
\end{abstract}

\paragraph{Keywords:} term deduction system, tree automata, graphical representation structural, proof theory, directed hypergraph, refinement system

\section{Introduction}

Structural proof theory \cite{Negri_vonPlato:2001} is a subfield of proof theory which focuses on the properties of derivations (aka. proofs) rather than on the derivable terms. For example, research in display calculi \cite{Greco_etal_LE:2018} is interested in structural rules and the shapes they exhibit in view of establishing meta-theorems, like cut-elimination.

Applications of such studies include improvements of methods used in proof-theoretic semantics, like categorial grammars used in computational linguistics \cite{Moortgat_Moot:2013}. The structure of the derivation have been shown to disambiguate structural meanings of natural language sentences, like quantifier scopes or subject vs. object relativization \cite{Correia_Stoof_Moortgat_spin:2004}.

Research in structural proof theory, but also in other fields involving formal methods, may lead to considering large deduction systems, containing several dozens of rules (e.g. 68 rules found in \cite{Greco_etal_focusing_display:2021}). Keeping track of all possible combinations of theses rules is a hard problem. This issue is particularly critical at the design phase, when trying to come up with a deduction system (aka. a calculus) which meets some requirements.

However, we do not just want to test \textit{whether} a calculus has the expected specification, but to know \textit{why} and \textit{how} it does or does not. The desiderata of calculus designers often revolves around intuitions about connectives and rules, e.g. ``What happens if I add or remove this rule?''.

The combinatorics of rules also brings a challenge at proof phase, when trying to demonstrate properties about a calculus. Many theorems on calculi still make use of case disjunction. Such a strategy becomes difficult and fastidious as the size of the system increases. There is a need to get a larger picture of calculi, to get new insights about them.

Approaches based on graphical languages, like proof nets or string diagrams, turned out to be of great use to give visual intuitions. Nevertheless, they often focus on a single derivation and divert from the very structure of derivation trees.

\subsection{Proposal}

In this article, we suggest to create a graphical representation of a calculus, or more generally, of any term deduction system. This representation is a sort of graph, where vertices are sets of terms and edges are rules. We call this representation a Proof Tree Graph (PTG).

As rules can have multiple hypotheses, a PTG is a directed hypergraph. Moreover, to relate sets of terms having a non-empty intersection, we allow distinguished edges.

The goal of a PTG is to give visual intuitions about the relationships between rules by linking the hypotheses and the conclusions of these rules. This way, it appears clearly how certain rules can follow other rules. Thus, a PTG illustrates the whole system, and not a particular derivation.

We also introduce a related notion of tree automata called Proof Tree Automata (PTA), which PTGs are the graphical representation. A PTA is a tree automaton augmented with two relations aiming at controlling that, while parsing a proof tree, hypothesis terms and conclusion terms are correctly related.


A forward proof-search in a given deduction system corresponds to a bottom-up run in a PTA of that system, and therefore to a graph (hyper)walk in the associated PTG.

Using automata and graphs is an open door to topological methods for term deduction. Even if we do not expand on this here, one goal of PTA and PTGs is to provide a tool with which we can translate properties expressed on sets of derivation trees into properties expressed on automaton runs or graph walks.

\subsection{Example}

To see how a graphical representation can really help intuition, let us design an example where it turns out to be useful.

We use three sorts $\sorta$, $\sortb$ and $\sortc$, and $a$ is an atomic symbol of $\sorta$. The sets of well-formed terms are given by:

\begin{equation}\label{eq:terms-ex}
\begin{array}{rrl}
\sorta \ni & u, v ::= & r(s) ~~|~~ f(s) ~~|~~ a \\
\sortb \ni & s, t ::= & l(u) ~~|~~ g(u) \\
\sortc \ni & h    ::= & u \vdaa v ~~|~~ u \vdab t ~~|~~ s \vdbb t
\end{array}
\end{equation}
where turnstile function symbols ($\vdaa$, $\vdab$ and $\vdbb$) are taken infix.

Let $\K$ denote the following term deduction system :

\begin{equation}\label{eq:calculus-ex}
\begin{array}{ccc}
\AXC{}
\RL{(Ax)}
\UIC{$u \vdaa u$}
\DP
~~ & ~~
\AXC{$u \vdaa r(t)$}
\RL{(Ad)}
\UIC{$l(u) \vdbb t$}
\DP
~~ & ~~
\AXC{$l(u) \vdbb t$}
\RL{(Ad')}
\UIC{$u \vdaa r(t)$}
\DP
\\[5mm]
 &
\AXC{$u \vdaa v$}
\RL{(g)}
\UIC{$u \vdab g(v)$}
\DP
&
\AXC{$u \vdab g(v)$}
\RL{(g')}
\UIC{$u \vdaa v$}
\DP
\end{array}
\end{equation}

$\K$ is clearly design as a multi-sorted sequent calculus. Rules (Ad) and (Ad') obviously refer to the adjunction (i.e. galois connection) $l \dashv r$. Similarly, rule ($g$') is the inverse rule of rule ($g$).\footnote{Actually, rule ($g$) is inspired by the shift up $\uparrow$.}

Let's call $\L_{\vda}(\K)$ the set of derivable terms of root connective $\vdaa$.

$\K$ has the following peculiar property : every term of $\L_{\vda}(\K)$ is of the form $u \vdaa u$. In other words, rules (Ad), (Ad'), ($g$) and ($g$') do not influence $\L_{\vda}(\K)$, neither individually, nor all together.

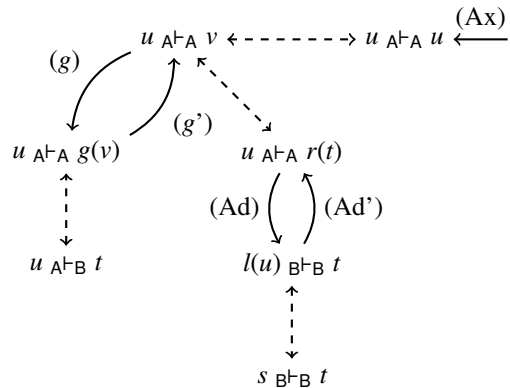
\begin{figure}[ht]
	\centering
\begin{tikzpicture}[thick,y=1.5cm,x=1.5cm]
\draw (0,0) node(uv){$u \vdaa v$};
\draw (2,0) node(uu){$u \vdaa u$};
\draw (-1,-1) node(ugv){$u \vdaa g(v)$};
\draw (1,-1) node(urt){$u \vdaa r(t)$};
\draw (1,-2) node(lut){$l(u) \vdbb t$};
\draw (-1,-2) node(ut){$u \vdab t$};
\draw (1,-3) node(st){$s \vdbb t$};
\draw (3,0) node(i){};
\draw[->] (uv) edge[bend right] node[above left]{($g$)} (ugv)
	(ugv) edge[bend right] node[below right]{($g$')} (uv)
	(urt) edge[bend right] node[left]{(Ad)} (lut)
	(lut) edge[bend right] node[right]{(Ad')} (urt)
	(i) edge node[above]{(Ax)} (uu)
	;
\draw[<->, dashed] (uv) edge (uu) edge (urt)
	(ut) edge (ugv)
	(lut) edge (st)
	;
\end{tikzpicture}
\caption{Proof tree graph $\G$ of $\K$. Vertices are terms with meta-variables. Full edges represent rules. Dashed edges connect vertices which share common instance terms.}
\label{fig:pta-ex}
\end{figure}

To visualize this easily, we can have a look at a proof tree graph $\G$ of $\K$, drawn in Figure \ref{fig:pta-ex}.

In $\G$, all hypotheses and conclusions appearing in \eqref{eq:calculus-ex} are vertices. We also added vertices corresponding to terms with meta-variables present in \eqref{eq:terms-ex}. Inverse rules are depicted as edges with swapped targets and sources. There is a dashed edge between $u \vdaa v$ and $u \vdaa r(t)$ because there is at least one term (e.g. $a \vdaa r(l(a))$) which is an instance of both. Same for the other dashed edges.

A derivation of $\K$ corresponds to a walk on $\G$. Given the walk $u \vdaa u \xrightarrow{\text{(g)}} u \vdab g(v)$, the only way to go back to $u \vdaa u$ is to take edge ($g$'), which cancels rule ($g$). As no other rule allows us to get to vertex $u \vdab g(v)$, $\G$ gives the good intuition that neither ($g$) nor ($g$') influences $\L_{\vda}(\K)$. We could summarize this subgraph shape by saying that ($g$), $u \vdab g(v)$ and ($g$') form a \textit{cul-de-sac} (i.e. a dead-end). This is an example of property on derivations which has a topological counterpart. This works similarly for rules (Ad) and (Ad').

Now suppose that we want to add rule \eqref{eq:extra-rule-ex} and assess its influence on $\L_{\vda}(\K)$.
\begin{equation}\label{eq:extra-rule-ex}
\AXC{$s \vdbb t$}
\RL{($f$)}
\UIC{$f(s) \vdab t$}
\DP
\end{equation}

Looking at $\K$ alone, it might be hard to have a quick idea of the possibilities added by ($f$). But adding the related edges on $\G$ (see Figure \ref{fig:pta-ex2}), we immediately see that the pattern of the whole term deduction system changes. The presence of ($f$) creates a loop which could enable non-trivial terms to be derived.

\begin{figure}[ht]
	\centering
\begin{tikzpicture}[thick,y=1.5cm,x=1.5cm]
\draw (0,0) node(uv){$u \vdaa v$};
\draw (2,0) node(uu){$u \vdaa u$};
\draw (-1,-1) node(ugv){$u \vdaa g(v)$};
\draw (1,-1) node(urt){$u \vdaa r(t)$};
\draw (1,-2) node(lut){$l(u) \vdbb t$};
\draw (-1,-2) node(ut){$u \vdab t$};
\draw (1,-3) node(st){$s \vdbb t$};
\draw (-1,-3) node(fst){$f(s) \vdab t$};
\draw (3,0) node(i){};
\draw[->] (uv) edge[bend right] node[above left]{($g$)} (ugv)
	(ugv) edge[bend right] node[below right]{($g$')} (uv)
	(urt) edge[bend right] node[left]{(Ad)} (lut)
	(lut) edge[bend right] node[right]{(Ad')} (urt)
	(i) edge node[above]{(Ax)} (uu)
	(st) edge node[above]{($f$)} (fst)
	;
\draw[<->, dashed] (uv) edge (uu) edge (urt)
	(ut) edge (ugv) edge (fst)
	(lut) edge (st)
	;
\end{tikzpicture}
\caption{Proof tree graph $\G'$ of $\K \cup \{$($f$)$\}$.}
\label{fig:pta-ex2}
\end{figure}

Indeed, thanks to ($f$), the following non-trivial term of $\L_{\vda}(\K)$ is derivable.
\begin{equation}
\AXC{}
\RL{(Ax)}
\UIC{$r(g(a)) \vdaa r(g(a))$}
\RL{(Ad)}
\UIC{$l(r(g(a))) \vdbb g(a)$}
\RL{($f$)}
\UIC{$f(l(r(g(a)))) \vdab g(a)$}
\RL{($g$')}
\UIC{$f(l(r(g(a)))) \vdaa a$}
\DP
\end{equation}

Therefore, proof tree graphs can be a useful tool helping to develop calculi and other formal systems.

\subsection{Outline}

We begin with a recall of definitions about term deduction systems in section \ref{sec:tds} to set a clear basis on the core notions. We define a version of tree automata on which PTA are based in section \ref{sec:tree-automata}. Section \ref{sec:graph} follows with the graphical representation. Proof tree automata and proof tree graphs are presented in section \ref{sec:pta}, as well as some simple properties that usually hold. We provide in section \ref{sec:comparison} a comparison of proof tree graphs with other graphical languages, namely string diagrams proof nets. Finally, we express PTA as refinement systems, revealing that PTA are traditional tree automata parametrized by a calculus.

\section{Term deduction systems}\label{sec:tds}

This section summarizes basic notions about term deduction systems (aka. calculi) and introduces the notations of this article.

Given a set $X$, $X^*$ is the set of words (i.e. finite sequences) on $X$. The empty word is $\varepsilon$ and $X^+$ stand for $X^* \setminus \set{\varepsilon}$. Concatenation of words $u$ and $v$ is $uv$. If $w \in X^*$, $w_i$ is its $i^{\text{th}}$ letter (beginning at $1$) and $\size{w}$ is the length of $w$.

If $n, m \in \nats$, $\interv{n,m}$ is the set of integers $k$ such that $n \leq k \leq m$.

\subsection{Signatures and terms}

We consider the general case of multi-sorted signatures without binding connectives.

\begin{Definition}\label{def:signature}
A \textbf{signature} $\Sign = (\S, \Sigma, \src, \trg)$ is given by
\begin{itemize}
    \item a non-empty finite set $\S$ which elements are called sorts
    \item a non-empty set $\Sigma$ which elements are called connectives (or function symbols)
    \item a source function $\src: \Sigma \to \S^*$
    \item a target function $\trg: \Sigma \to \S$
\end{itemize}

We write $\Sigma_{s_1...s_n}$ the set of connectives $f \in \Sigma$ such that $\src(f) = s_1...s_n$ and $\Sigma^s$ the set of connectives $f$ such that $\trg(f) = s$. As expected, $\Sigma^s_{s_1...s_n}$ stands for $\Sigma^s \cap \Sigma_{s_1...s_n}$.

We say that a connective $f$ is of arity $n = \size{\src(f)}$ and we write $\Sigma_n$ the set of connectives of arity $n$.
\end{Definition}

If not mentioned explicitly, a signature is supposed to be finite, i.e. $\Sigma$ is finite.

We use the definition of trees exposed in \cite{TATA:2008}.

\begin{Definition}
A finite rooted labelled tree $t = (\tau, L, \lambda)$ is given by a set $\tau \subseteq \nats^*$ of elements called nodes, a set $L$ and a labelling function $\lambda : \tau \to L$, verifying
\begin{enumerate}
    \item If $\nu \in \tau$, then all prefixes of $\nu$ belong to $\tau$
    \item If $\nu i \in \tau$ and $j \leq i$, then $\nu j \in \tau$
\end{enumerate}

The empty word $\varepsilon$ is called the root. A node $\nu \in \tau$ is called a leaf if $\nu 0 \not \in \tau$, otherwise it is called an internal node. The daughters of a node $\nu \in \tau$ is the set of $\nu i$ which belong to $\tau$ for some $i \in \nats$. We call arity of $\nu$ the cardinal of this set. 
\end{Definition}

We may avoid mentioning the codomain $L$ of $\lambda$ when it is clear from context.

\begin{Definition}\label{def:term}
A \textbf{term} $t$ on a signature $\Sign = (\S, \Sigma, \src, \trg)$ is a non-empty finite rooted labelled tree $t =(\tau, \Sigma, \lambda)$ satisfying, for all node $\nu \in \tau$ of daughters $\nu_1,...,\nu_n$ ($n \in \nats$),
\begin{equation}
    \src(\lambda(\nu)) = \trg(\lambda(\nu_1))...\trg(\lambda(\nu_n))
\end{equation}

If $\rho$ is the root of $\tau$, we write $\trg(t) = \trg(\lambda(\rho))$ the sort of the term.

We write $\T(\Sign)$ the set of terms on $\Sign$.
\end{Definition}

\begin{Remark}
We might as well see a term $t \in \T(\Sign)$ as defined by mutual induction:
\begin{equation}
    \T(\Sign)^s \ni t_s ::= f(t_{s_1}, ..., t_{s_n}),~ f \in \Sigma^s_{s_1...s_n}
\end{equation}
\end{Remark}

\begin{Definition}
A \textbf{variable set} $\Varia = (\S, \V, \trg)$ is given by
\begin{itemize}
    \item a finite set $\S$ which elements are called sorts
    \item a countably infinite set $\V$ which elements are called variables
    \item a target function $\trg : \V \to \S$
\end{itemize}
\end{Definition}

\begin{Definition}
If $\Sign = (\S, \Sigma, \src, \trg)$ is a signature and $\Varia = (\S, \V, \trg')$ a variable set on the same set of sorts and such that $\Sigma \cap \V = \emptyset$, we write $\T(\Sign, \Varia)$ the set of terms on $(\S, \Sigma \cup \V, \src + \src', \trg + \trg')$
\footnote{The notation $+$ means here the union of functions on disjoints domains.}
, where $\src'(x) = \varepsilon$ for all $x \in \V$.
\end{Definition}

To declare signatures, we may simply write $f\t{s_1...s_n, s}$ to mean $\src(f) = s_1...s_n$ and $\trg(f) = s$, and similarly for variable sets.

\begin{Definition}
Set $\Sign$ a signature and $\Varia$ a variable set on the same sort set $\S$.

A \textbf{substitution} is a partial function $\sigma : \V \nrightarrow \T(\Sigma, \V)$ such that if $\sigma(x) = t$, then $\trg(x) = \trg(t)$.

The application of $\sigma$ to a term $t \in \T(\Sigma, \V)$ is a term $t \sigma \in \T(\Sigma, \V)$ defined by induction

\begin{equation}
\begin{array}{rll}
     x \sigma & = \sigma(x) & \text{if $x \in \dom \sigma$} \\
     x \sigma & = x & \text{if $x \not\in \dom \sigma$} \\
     f(t_1,...,t_n) \sigma & = f(t_1 \sigma,..., t_n \sigma)
\end{array}
\end{equation}

We write $\subst{\repl{x_1}{t_1},..., \repl{x_n}{t_n}}$ for the substitution $\sigma : x_i \mapsto t_i$ of domain $\set{x_1,...x_n}$.
\end{Definition}

%
%

\begin{Example}
	In this paper we use implicational logic $\ImpL$ as illustration example, because it is small but exposes interesting basic properties. We set a finite set $\Att$ of atomic types. The signature $\Sign_{\ImpL}$ is given by
	\begin{equation}
	\begin{array}{rrl}
	\mathsf{Form} \ni & \phi, \psi ::= & a \in \Att ~|~ \phi \to \psi \\
	\mathsf{Cont} \ni & \Gamma, \Delta ::= & \phi ~|~ \Gamma, \Delta \\
	\mathsf{Seq} \ni & \chi ::= & \Gamma \vdash \phi \\
	\end{array}
	\end{equation}
	
	We do not use variables. The letters $\phi, \psi, \Gamma$ and $\Delta$ (potentially with subscripts) are used as meta-variables (see appendix \ref{sec:schematic}).
	
	Note that we actually implicitly use associativity of the context connective ``$,$''. This departs from our settings, which does not allow terms up to equations. But this provides a more common baseline than non-associative contexts.
\end{Example}

\subsection{Term deduction systems}

A term deduction system is a formal system where every statement is derived from recursively applying rules on axiomatic statements.

\begin{Definition}\label{def:rule}
Set $\Sign$ a signature and $\Varia$ a variable set on the same sort set $\S$.
A sort-consistent \textbf{rule} is a set $R \subseteq \T(\Sign, \Varia)^* \times \T(\Sign, \Varia)$ which elements are called rule instances. We additionally assume that there exists $s_1,...,s_n, s \in \S$ such that for every $(u, t) \in R$, $\trg(t) = s$ and $u = t_1...t_n$ with $\trg(t_i) = s_i$ for all $i \in \interv{1,n}$. We may write as well $\trg(R) = s$ and $\src(R) = s_1...s_n$.
\end{Definition}

In the following, all rules are supposed to be sort-consistent.

\begin{Definition}\label{def:sort-consistent}

Given a rule $R$, we can define the following notation:
\begin{equation}
\begin{array}{rll}
    \dom R = & \cset{t_1...t_n}{\exists t, (t_1...t_n, t) \in R} \\
     \dom[i] R = & \cset{t_i}{\exists t, t_1..., t_{i-1}, t_{i+1},... t_n, (t_1...t_i...t_n, t) \in R} & \text{for } i \in \interv{1, n}\\
     \codom R = & \cset{t}{\exists t_1... t_n, (t_1...t_n, t) \in R}
\end{array}
\end{equation}

\end{Definition}

\begin{Definition}\label{def:tds}
A \textbf{term deduction system} or \textbf{calculus} (TDS) is a triplet $\K = (\Sign, \Varia, \R)$ where
\begin{itemize}
    \item $\Sign$ a signature
    \item $\Varia$ a variable set on the same sort set $\S$
    \item $\R$ is a set of rules on $\Sign$ and $\Varia$
\end{itemize}

We write $\R_n$ the set of rules of $\R$ of arity $n$, i.e. having $n$ hypotheses.
\end{Definition}

\begin{Definition}\label{def:derivation}
A \textbf{derivation} of a TDS is a term $(\pi, \lambda)$ on the infinite signature $(\S, \T(\Sign, \Varia) \times \R, \src, \trg)$ where we define $\trg((t,R)) = \trg(R)$ and $\src((t,R)) = \src(t)$ for every $(t, R)$, and such that
\begin{itemize}
    \item for every node $\nu \in \pi$ of daughters $\nu_1,...,\nu_n$, if $\lambda(\nu) = (t, R)$ and $\lambda(\nu_i) = (t_i, R_i)$ for all $i$, then $(t_1...t_n, t) \in R$
\end{itemize}

If $t$ is the root term of a derivation $\pi$, we say that $t$ is derivable in $\K$ and we write $\vdash_{\K} t$. We write $\L(\K)$ the language of $\K$, i.e. the set of terms $t$ such that $\vdash_{\K} t$. We write $\D(\K)$ the derivation language of $\K$, i.e. the set of derivations on $\K$.
\end{Definition}

\begin{Example}
	The calculus $\K_{\ImpL}$ of implicational logic is given by
\begin{equation}
\begin{array}{ccc}
\AXC{}
\RL{Ax.}
\UIC{$\phi \vdash \phi$}
\DP
&
\AXC{$\Delta, \phi \vdash \psi$}
\RL{$\to$ I.}
\UIC{$\Delta \vdash \phi \to \psi$}
\DP
&
\AXC{$\Delta \vdash \phi \to \psi$}
	\AXC{$\Gamma \vdash \phi$}
\RL{$\to$ E.}
\BIC{$\Delta, \Gamma \vdash \psi$}
\DP
\\[4mm]
\AXC{$\Gamma \vdash \phi$}
\RL{Weak.}
\UIC{$\Gamma, \psi \vdash \phi$}
\DP
&
\AXC{$\Delta, \phi, \phi \vdash \psi$}
\RL{Contr.}
\UIC{$\Delta, \phi \vdash \psi$}
\DP
&
\AXC{$\Gamma, \phi_1, \phi_2, \Delta \vdash \psi$}
\RL{Exch.}
\UIC{$\Gamma, \phi_2, \phi_1, \Delta \vdash \psi$}
\DP
\end{array}
\end{equation}
\end{Example}

\section{Controlling tree automata}\label{sec:tree-automata}





In this section we design an extension of tree automata which can run on proof trees. Proof tree automata are defined in section \ref{sec:pta} as special cases of this extended automata dependent on a calculus.

Derivations cannot be simply put into a traditional tree automaton \cite{TATA:2008} because nodes encapsulate complex information (i.e. terms) on which complex operations have to process. We suggest to enrich tree automata with two relations called controlling relations, which purpose is to ensure that these operations are processed correctly.

In view of proof trees, we assume that the set of node labels can be split into a cartesian product. One component $A$ is finite and accounts for the signature properties. The other component $\Phi$ may be infinite.

In view of proof tree graphs, we also allow $\varepsilon$-transitions.

\begin{Definition}\label{def:ncta}
A \textbf{non-deterministic controlling tree automata with $\varepsilon$-transitions} (\NCTAe) is a tuple $\A = (A, \Phi, Q, \delta, \nabla, F, \delta\eps, \nabla\eps)$, where
\begin{itemize}
    \item $A = (\S^A, \Sigma^A, \src^A, \trg^A)$ is a finite signature and $\Sigma^A$ is called the alphabet
    \item $\Phi$ is a set of elements called instances
    \item $Q$ is a finite set of elements called states
    \item $\delta \subseteq \bigcup_{n \in \nats} Q^n \times \Sigma^A_n \times Q$ is the set of transitions
    \item $\nabla \subseteq \Phi^n \times \Sigma^A \times \Phi$ is the controlling relation
    \item $F \subseteq Q$ are the final states
    \item $\delta\eps \subseteq Q \times Q$ is the set of $\varepsilon$-transitions, 
    \item $\nabla\eps \subseteq \Phi \times Q$ is the controlling relation for $\varepsilon$-transitions, also called $\varepsilon$-controlling relation
\end{itemize}
\end{Definition}

\begin{Remark}
The controlling relation $\nabla$ does not have to be viewed has a transition set. It rather verifies if the instance $t \in \Phi$ of as node label is consistent with its $\Sigma^A$-component read by the transition and the instances $t_1,...t_n$ of its daughters. Elements of $\Phi$ are called instances because we could interpret the $\varepsilon$-controlling set $\nabla\eps$ as the relation ``is an instance of'', i.e. states $q \in Q$ abstract over elements of $\Phi$.
\end{Remark}






A run of a tree automaton $\A$ on tree $D$ is a labelling of the nodes of $D$ by states of $\A$. Here we map nodes to non-empty words of states, to account for $\varepsilon$-transitions.

\begin{Definition}\label{def:run-NCTAe}
A \textbf{run} $\gamma$ of a \NCTAe $\A = (A, \Phi, Q, \delta,  \nabla, F,\delta\eps, \nabla\eps)$ on a term $D = (\pi, \Phi \times \Sigma^A, \lambda)$ is a labelling $\gamma : \pi \to Q^+$ such that for every node $\nu \in \pi$ labelled by $(t, a)$:
\begin{enumerate}
    \item\label{it:run-NTCAe-delta} if the daughters of $\nu$ are $\nu_1,..., \nu_n$ then $(\gamma(\nu_1)_{m_1}... \gamma(\nu_n)_{m_n}, a, \gamma(\nu)_1) \in \delta$, where $m_j = \size{\gamma(\nu_j)}$ \hfill (transitions)
    
    \item\label{it:run-NTCAe-nabla} by noting $ \lambda(\nu_j) = (t_j, a_j)$ for $1 \leq j \leq n$, then $(t_1... t_n, a, t) \in \nabla$ \hfill (control)
    
    \item\label{it:run-NTCAe-eps} for every $1 \leq i < \size{\gamma(\nu)} - 1$, $(\gamma(\nu)_i, \gamma(\nu)_{i+1}) \in \delta\eps$, \hfill ($\varepsilon$-transitions)
    
    \item\label{it:run-NTCAe-nabla-eps} and $(t, \gamma(\nu)_{i+1}) \in \nabla\eps$ \hfill ($\varepsilon$-control)
\end{enumerate}

A run on $D$ is accepting if $\gamma(\varepsilon) \in F$. The language of $\A$ is $\L(\A)$ the set of terms on which there is an accepting run of $\A$.
\end{Definition}

\begin{Example}
	Let's consider an example outside the realm of logic. Set $A$ the signature with only one sort and the connectives $0$ of arity $0$, $\Incr$ of arity $1$ and $\Add$ of arity $2$.
	We define $Q = \set{\zero, \even, \odd}$ with $\zero = \set{0}$, $\even = 2\nats$ and $\odd = 2\nats +1$. Let $\A_{\Arthm}$ be the \NCTAe $(A, \nats, Q, \delta, \nabla, Q, \delta\eps, \nabla\eps)$ with
	\begin{equation}
	\begin{array}{rl}
	\delta = & \{(\varepsilon, 0, \zero), (\even, \Incr, \odd), (\even, \Incr, \odd), (\even~\even, \Add, \even), (\odd~\odd, \Add, \even), \\ & ~~ (\odd~\even, \Add, \odd), (\even~\odd, \Add, \odd)\} \\
	\nabla = & \set{(\varepsilon, 0, 0)} \cup \cset{(n, \Incr, n+1)}{n \in \nats} \cup \cset{(n~m, \Add, n+m)}{n,m \in \nats} \\
	\delta\eps = & \set{(\zero, \even)} \\
	\nabla\eps = & \cset{(n, \even)}{n \in \even}
	\end{array}
	\end{equation}
	
	$\A_{\Arthm}$ is computing the result of a tree of basic arithmetic operations while keeping track of the parity as states. $A$ is the set of operations and $\Phi = \nats$ is the actual elements on which the operation is performed. The set of transitions $\delta$ indicates how operations change the state. The controlling relation $\nabla$ is actually giving the semantics of the operation on elements. The set of $\varepsilon$-transitions $\delta\eps$ indicates how the automaton can jump from one state to another without reading an operation. The $\varepsilon$-controlling relation $\nabla\eps$ rules whether, knowing the element, it is allows to use an $\varepsilon$-transition. Note that this always only depend on the target state.
	
	An example tree belonging to $\L(\A_{\Arthm})$ is given in Fig.~\ref{fig:tree-arthm}.
\end{Example}

\subsection{Consistency}

A relevant property about \NCTAe is consistency: the fact that every transition target state is consistent (w.r.t. $\nabla\eps$) with the set of image instances of this state given by $\nabla$.

\begin{Definition}\label{def:pta}
	A \NCTAe $\A = (A, \Phi, Q, \delta,  \nabla, F,\delta\eps, \nabla\eps)$ is \textbf{consistent} if
	for every transition $(q_1...q_n, a, q) \in \delta$, and instances $t, t_1,..., t_n$:
	\begin{equation}
	\text{if } (t_1...t_n, R, t) \in \nabla \text{ and } (t_i, q_i) \in \nabla\eps \text{ for all } i \in \interv{1,n} \text{, then } (t, q) \in \nabla\eps
	\end{equation}
\end{Definition} 

Consistency is a very desirable property because it asserts that different parts of a \NCTAe have some elementary coherence.

\begin{Example}
	$\A_{\Arthm}$ is consistent.
	
	If we define $\A_{\Arthm}^2$ as $\A_{\Arthm}$ with an additional transition $ \xrightarrow{0} \odd$, then $\A_{\Arthm}^2$ is not consistent because $0$ is not odd. However, this does not affect the language, i.e. $\L(\A_{\Arthm}) = \L(\A_{\Arthm}^2)$.
\end{Example}


%
%
%


\section{Graphical representation}\label{sec:graph}

Here we present a graphical language to represent non-deterministic controlling tree automata with $\varepsilon$-transitions, and we expose graphical conventions. We first introduce an extended notion of directed hypergraph to account for the various transitions of such automata. Definitions are adapted from \cite[chap. 6]{Bretto:2013}.

\begin{Definition}
A \textbf{labelled directed hypergraph} (DHG) is a tuple $G = (V, E, L)$ where $V$ is the set of vertices, $L$ the set of labels and $E \subseteq V^* \times L \times V^*$ is the set of edges.

If $w \in V^*$, we set the following notation: $E(w) = \cset{u \in V^*}{\exists l \in L, (w,l,u) \in E}$, $E_n = E \cap (V^n \times L \times V^*)$ for any $n \in \nats$.
\end{Definition}

In the following, we only consider DHGs where edges have \textit{exactly one target vertex}, i.e. ${E \subseteq V^* \times L \times V}$.

As we consider sorted signatures, we also want to consider that graph nodes have a sort whenever labels are the function of a signature. In proof tree graphs, labels are taken to be rules, which by definition have a fixed sort-arity.

\begin{Definition}
If $\Sign= (\S, \Sigma, \src, \trg)$ is a signature, then we say that the DHG $G = (V, E, \Sigma)$ is \textbf{typed by} $\Sign$ if there is a function $h : V \to \S$ such that the edge source and target $h$-labels are consistent with the sort-arity of the $\Sigma$-labels, i.e.:

\begin{itemize}
    \item for all $(w, f, v) \in E_n$ we have $\trg(f) = h(v)$, $f \in \Sigma_n$, $\size{w} = n$ and $\src(f) = h(w_1)...h(w_n)$, for some $n$
\end{itemize}
\end{Definition}


\begin{Definition}
A \textbf{DHG with dashed unary edges} (\DHGd) is a tuple $G = (V, E, L, E_d)$ where $(V, E \cup E_d, L)$ is a DHG and $E_d \subseteq V \times V$ is a set of label-less additional unary edges.

We represent a $E_d$-edge $(u, v)$ by a dashed arrow $u \darrow v$.
\end{Definition}

The usual notion of walk in a directed graph extends to hyperwalks in a directed hypergraph with dashed unary edges.

\begin{Definition}\label{def:hyperwalk}
A \textbf{hyperwalk} in a \DHGd $G = (V, E, L, E_d)$ is a finite rooted labelled tree $H = (\tau, (E \times V) (E_d \times V)^*, \lambda)$ such that for every node $\nu \in \tau$ of daughters $\nu_1,..., \nu_n$, if $\lambda(\nu) = (e_1, u_1)...(e_m, u_m)$ and $\lambda(\nu_i) = w(e^i, u^i)$ for all $i$, then
\begin{itemize}
	\item  $e_1 = (u^1...u^n, l, u_1)$ for some $l$ \hfill ($n$-ary edges)
	\item for all $1 < j \leq m$, $e_j = (u_{j-1}, u_j)$ \hfill (unary edges)
\end{itemize}
\end{Definition}

%

\subsection{Links between \texorpdfstring{\NCTAe}{NCTAe} and \texorpdfstring{\DHGd}{DHGd}}\label{sec:link}

As explained above, we want to represent our extended tree automata by hypergraphs the same way string automata are represented on a paper by a labelled directed graph.

\begin{Definition}\label{def:underlying-graph}
Given a \NCTAe $\A = (A, \Phi, Q, \delta, \nabla, F, \delta\eps, \nabla\eps)$, its \textbf{underlying graph} is defined as the \DHGd ${G = (Q, \delta, \Sigma^A, \delta\eps)}$.
\end{Definition}

\begin{Remark}
	By definition \ref{def:ncta}, $A$ is a finite signature. Therefore, the underlying graph of $\A$ is typed by $A$.
\end{Remark}

Accepting states and controlling relations are not represented on the underlying graph. Therefore, there is some information loss.

The reverse transformation of definition \ref{def:underlying-graph} is the following.

\begin{Definition}\label{def:aut-repr-by}
Fix a \DHGd $G = (V, E, \Sigma, E_d)$ typed by $\Sign = (\S, \Sigma, \src, \trg)$. We define the \textbf{automaton represented by $G$} as the \NCTAe $\A = (\Sign, \T(\Sign), V, E, \nabla, V, E_d, \nabla\eps)$, where 

\begin{equation}
\begin{array}{rcl}
\nabla & = & \cset{(D_1...D_n, f(D_1,...,D_n), f(D_1,...,D_n))}{f \in \Sigma_{s_1...s_n}, D_i \in \T(\Sign)^{s_i}} \\
\nabla\eps & = & \T(\Sign) \times V
\end{array}
\end{equation}
\end{Definition}

The following property shows that taking the automaton represented by a typed \DHGd does not lose any information.

\begin{Proposition}
The underlying graph of the automaton represented by a typed \DHGd $G$ is $G$.
\end{Proposition}

\begin{proof}
	Straightforward by applying the definitions.
\end{proof}

\begin{Example}
The underlying graph $\G_{\Arthm}$ of example $\A_{\Arthm}$ is given in Fig.~\ref{fig:graph-arthm}. The transitions $\even ~ \odd \xrightarrow{\Add} \odd$ and $\odd ~ \even \xrightarrow{\Add} \odd$ are not distinguished to increase readability.
\end{Example}

\begin{figure}[ht]
\begin{minipage}{.45\textwidth}
\centering
\begin{tikzpicture}
\Tree [.$(3,\Add)$ [.$(1,\Incr)$ $(0,0)$ ]
		[.$(2,\Add)$ [.$(2,\Incr)$ [.$(1,\Incr)$ $(0,0)$ ] ] $(0,0)$ ]
	]
\end{tikzpicture}
\captionof{figure}{Tree belonging to the language of $\A_{\Arthm}$}
\label{fig:tree-arthm}
\end{minipage}
\hfill
\begin{minipage}{.5\textwidth}
	\centering
\begin{tikzpicture}[thick,>=latex,outer sep=1mm]
\draw[circle] (0.5,-1) node[draw](0){$\zero$};
\draw[circle] (2,0) node[draw](e){$\even$};
\draw[circle] (4,0) node[draw](o){$\odd$};
\draw (1.5,.5) node[inner sep=0,outer sep=0,minimum size=0](i1){};
\draw (4.5,-.5) node[inner sep=0,outer sep=0,minimum size=0](i2){};
\draw (4.5,.8) node[inner sep=0,outer sep=0,minimum size=0](i3){};
\draw (-1,-1) node(i){};
\draw[->] (i) edge node[above]{$0$} (0) ;
\draw[dashed,->] (0) edge (e) ;
\draw[->] (e) edge[bend left] node[above]{$\Incr$} (o) ;
\draw (e) edge[out=145] (i1) edge[out=110] (i1)
	edge[out=90,bend left=80] (i3) ;
\draw[->] (o) edge[bend left] node[above]{$\Incr$} (e) ;
\draw (o) edge[out=90] (i3)
	edge[out=-40,bend left] (i2) edge[bend right,out=-20] (i2) ;
\draw[->] (i1) edge[bend right=100,in=-140] node[above left]{$\Add$} (e)
(i2) edge[bend left=70] node[above]{$\Add$} (e)
(i3) edge[bend left] node[right]{$\Add$} (o) ;
\end{tikzpicture}
\captionof{figure}{underlying graph $\G_{\Arthm}$ of $\A_{\Arthm}$}
\label{fig:graph-arthm}
\end{minipage}
\end{figure}


%
%

A PTG does not carry control relations. This is reflected in the definition of a hyperwalk, where we are allowed to use labelled edges and dashed edges without restrictions. This formally manifests in definition \ref{def:hyperwalk}, where conditions ($n$-ary edges) and (unary edges) respectively correspond to conditions (transitions) and ($\varepsilon$-transitions) of definition \ref{def:run-NCTAe}, but conditions (control) and ($\varepsilon$-control) are missing.

We call correct the hyperwalks which correspond to actual runs in the PTA.

\begin{Definition}\label{def:correction}
Set $\A$ a \NCTAe and $\G$ the underlying graph of $\A$.	

A hyperwalk $H = (\pi, \lambda_H)$ in $\G$ is called \textbf{correct} w.r.t. $\A$ if there exists a proof tree $D = (\pi, \lambda_D)$ and a run $\gamma : \pi \to Q^+$ on $D$ in $\A$ such that, for all node $\nu \in \pi$ of daughters $\nu_1,..., \nu_n$, if $\lambda_D(\nu) = (t, R)$ then
\begin{equation}\label{eq:correction}
\begin{array}{rrcl}
& \lambda_H(\nu) & = & (d, \gamma(\nu)_1)(d_1, \gamma(\nu)_2)... (d_{m-1},\gamma(\nu)_m) \\
\text{with } & d & = & (\gamma(\nu_1)_{m_1}...\gamma(\nu_n)_{m_n}, R, \gamma(\nu)_1) \\
\text{and } & d_k & = & (\gamma(\nu)_k, \gamma(\nu)_{k+1})
\end{array}
\end{equation}
where $m_j = \size{\gamma(\nu_j)}$.
\end{Definition}

\section{Proof tree automata}\label{sec:pta}

In this section, we introduce proof tree automata (PTA) as spacial cases of \NCTAe. After the main definitions and some words about schematic PTA, we exhibit a completeness criterion (section \ref{sec:completeness}) to relate the language of a PTA to the derivation language of its corresponding calculus. In section \ref{sec:other-properties}, we define canonical PTA and we investigate another properties of PTA called totality, which arises with modular calculi.

In what follows, we fix a calculus $\K = (\Sign, \Varia, \R)$ on the signature $\Sign = (\S, \Sigma, \src, \trg)$.

\begin{Definition}\label{def:ta}
A \textbf{proof tree automaton (PTA) on $\K$} is an \NCTAe $\A = (A, \Phi, Q, \delta, \nabla, F, \delta\eps, \nabla\eps)$ where we require
\begin{itemize}
    \item $A = (\S, \R, \src^{\R}, \trg^{\R})$, sorted as in definition \ref{def:rule} \hfill (transition labels are rules)
    \item $\Phi = \T(\Sign, \Varia)$ \hfill (instances are terms)
    \item $Q \subseteq_f \wp(\Phi) \setminus \set{\emptyset}$ \hfill (states are non-empty sets of terms)
    \item $(t_1...t_n, R, t) \in \nabla$ iff $(t_1...t_n, t) \in R$ \hfill (control is given by rules)
    \item $F = Q$ \hfill (all states are accepting)
    \item $(t, q) \in \nabla\eps$ iff $t \in q$ \hfill ($\epsilon$-control is membership)
\end{itemize}

The symbol $\subseteq_f$ here means ``is a finite subset of''.
\end{Definition}

\begin{Remark}\label{rem:trans-and-dom}
	In a rule transition $(q_1...q_n, R, q)$, $q$ does not need to be the conclusion of $R$. This also holds for $q_1,..., q_n$ and the hypotheses of $R$.
\end{Remark}

\begin{Definition}\label{def:ptg}
A \textbf{proof tree graph (PTG) on $\K$} is the underlying graph of a PTA on $\K$.
\end{Definition}

\begin{Notation}
	In a PTA $\A = (A, \Phi, Q, \delta, \nabla, F, \delta\eps, \nabla\eps)$ on $\K$, as $A$, $\Phi$, $F$, $\nabla$ and $\nabla\eps$ are completely determined by $\K$, we may declare $\A$ with the following tuple: even $\A = (\K, Q, \delta, \delta\eps)$ to avoid redundancy. The symbol $\Phi$ may still be used to mean the set of terms on the signature of $\K$.
\end{Notation}

By construction, the language of a proof tree is always included in the derivation language of the calculus.

\begin{Proposition}\label{prop:soundness}
	For every proof tree automaton $\A$ on $\K$, $\L(\A) \subseteq \D(\K)$.
\end{Proposition}

\begin{proof}
	Set $\A = (\K, Q, \delta, \delta\eps)$ a PTA. Let $\gamma : \pi \to Q$ be an execution of $\A$ on a term $D = (\pi, \Phi \times \Sigma^A, \lambda)$. Let us show that $D$ is a derivation of $\K$ by induction on $\pi$.
	
	By definition \ref{def:derivation}, derivations are non-empty.
	
	\begin{itemize}
		\item Suppose $\pi = \set{\varepsilon}$ and $\lambda(\varepsilon) = (t, R)$. We note $\gamma(\varepsilon) = q_0...q_m$. We have $(\varepsilon, R, q_0) \in \delta$ and $(\varepsilon, R, t) \in \nabla)$ by definition \ref{def:run-NCTAe}. The latter gives us $(\varepsilon, t) \in R$ because $\A$ is a PTA. So $D \in \D(\K)$.
		\item Now suppose $D = (t, R) (D_1,..., D_n)$. We note $\gamma_{D_j}$ the sub-run on $D_j$, and $\gamma(\varepsilon) = q_0...q_m$. By induction hypothesis on every $D_j$, $D_j \in \D(\K)$. Let us write $\lambda(j) = (t_j, R_j)$ and $q_j' = \gamma_{D_j}(\varepsilon)_{n_j} = \gamma(j)_{n_j}$, where $n_j = \size{\gamma_{D_j}(\varepsilon)}$. Given that $(q_1',...,q_n', R, q_0) \in \delta$ and $(t_1...t_n, R, t) \in \nabla$ by definition \ref{def:run-NCTAe}, we have $(t_1...t_n, t) \in R$. So $D \in \D(\K)$. 
	\end{itemize}
\end{proof}

\begin{Remark}
	If $(q_1...q_n, a, q)$ is a transition in a PTA $\A$, consistency of $\A$ means that $q \subseteq \codom R$, and actually that $q$ intersects $\codom R$ because $q_1,..., q_n$ are non-empty.
\end{Remark}

\subsection{Examples}

\begin{figure}[ht]
	\centering
	\begin{tikzpicture}[thick]
	\draw (0,0) node(s){$\Gamma \vdash \phi$};
	\draw (-2,0) node(a){$\phi \vdash \phi$};
	\draw (0,2) node(w){$\Delta, \phi \vdash \psi$};
	\draw (-2,-1) node(e){$\Gamma, \phi_1, \phi_2, \Delta \vdash \psi$};
	\draw (0,-2) node(FL){$\Delta, \Gamma \vdash \psi$};
	\draw (3,-1) node(fl){$\Delta \vdash \phi \to \psi$};
	\draw (2, 1) node(c){$\Delta, \phi, \phi \vdash \psi$};
	\draw (-4,0) node(d){};
	\draw (1,-1) node(i)[inner sep=0,outer sep=0,minimum size=0]{};
	\draw (i) edge (s) edge (fl);
	\draw[->] (d) edge node[above]{Ax.} (a)
	(s) edge[bend left] node[left]{Weak.} (w)
	(c) edge node[above right]{Contr.} (w)
	(e) edge[loop below] node[below]{Exch.} (c)
	(w) edge[bend left=65] node[right]{$\to$ I.} (fl)
	(i) edge node[right]{$\to$ E.} (FL)
	;
	\draw[->,dashed] (a) edge (s)
	(s) edge[<->] (w) edge (c) edge[<->] (fl) edge[<->] (e)
	(FL) edge (s)
	;
	\end{tikzpicture}
	\caption{Consistent, complete and total PTA $\A_{\ImpL}$ for $\ImpL$}
	\label{fig:pta-ImpL}
\end{figure}
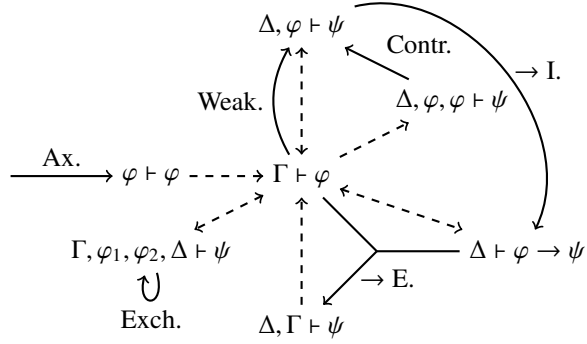

\begin{Example}
Figure \ref{fig:pta-ImpL} shows the PTG of a proof tree automata $\A_{\ImpL}$ on implicational logic $\ImpL$. Schematic terms $\term$ are used to represent sets of sequents $\inst{\term}$ (see appendix \ref{sec:schematic}).

Double headed dashed arrows simple stand for a dash arrow in both directions, to be less cumbersome.

Remark that for some cases (the target of Weak. and the target of Exch.), the schematic terms used in the rules are not the same as the ones given in $\A_{\ImpL}$, but have the same instance set.
\end{Example}

\begin{Example}
$\A_{\Arthm}$ can be seen as a PTA on the following term deduction system $\K_{\Arthm}$ with infinitely many rules:
\begin{equation}
\begin{array}{ccc}
\AXC{}
\RL{$0$}
\UIC{$0$}
\DP
&
\AXC{$n$}
\RL{$\Incr$}
\UIC{$n+1$}
\DP
&
\AXC{$n$}
	\AXC{$m$}
\RL{$\Add$}
\BIC{$n+m$}
\DP
\end{array}
\end{equation}
where $+$ is a meta-language symbol.
\end{Example}

\subsection{Completeness}\label{sec:completeness}

Now we are interested in a completeness criterion.

Intuitively, a PTA $\A$ is complete when, for every rule $R$ of the calculus, if the hypotheses of $R$ can be recognized by $\A$, then there exists a ``general'' transition from these hypotheses to a conclusion of $R$. By ``general'' transition we mean a hyperwalk which can be decomposed into $\varepsilon$-paths from the sources, and then one rule path labelled by $R$.

In the following definition, we explain what $\varepsilon$-path we consider.

\begin{Definition}
Let $\A = (\K, Q, \delta, \delta\eps)$ be a proof tree automaton, $q^0 \in Q$ a state and $t$ a term such that $t \in q^0$. We say that $q^0,...,q^{\ell}$ is a \textbf{$t-\varepsilon$-path} starting at $q^0$ in $\A$ if
\begin{equation}
\text{for all } 0 \leq i < \ell \text{, } (q^i, q^{i+1}) \in \delta\eps \text{ and } t \in q^{i+1}
\end{equation}

\begin{center}
	\begin{tikzpicture}[inner sep=0, thick, outer sep=1mm, label distance=0mm]
	\tikzset{nod/.style={draw,inner sep=.7mm,fill=black,circle}}
	\draw (0,0) node(11)[nod,label={90:$t \in q^0$}]{};
	\draw (1,0) node(12)[nod,label={90:$t \in q^1$}]{};
	\draw (4,0) node(1l)[nod,label={90:$t \in q^{\ell}$}]{};
	\draw (2,0) node(13){};\draw (3,0) node(14){};
	\draw (11) edge[dashed,->] node{} (12);
	\draw (12) edge[dashed,->] node{} (13);
	\draw (14) edge[dashed,->] node{} (1l);
	\draw (2.5,0) node{$...$};
	\end{tikzpicture}
\end{center}

We write $\varepsilon P_{\A}(t, q^0)$ the set of $t-\varepsilon$-paths starting at $q^0$ in $\A$.
\end{Definition}

\begin{Definition}
A \textbf{conclusion state} of $\A$ is a state $q \in Q$ such that there exists $R \in A$ and $q_1...q_n$ such that $(q_1...q_n, R, q) \in \delta$.
\end{Definition}

\begin{Definition}\label{def:completeness}
A proof tree automaton $\A = (\K, Q, \delta, \delta\eps)$ is \textbf{complete} if for every conclusion states $q_1,...,q_n$, terms $t, t_1,..., t_n$ and rule $R$ such that $t_i \in q_i$ for all $i \in \interv{1,n}$ and $(t_1...t_n, R, t) \in \nabla$:
\begin{equation}
\text{there exists } q \in Q \text{, $\varepsilon$-paths } q_j^0,..., q_j^{\ell_j} \in \varepsilon P_{\A}(t_j, q_j) \text{ and a transition } (q_1^{\ell_1},...,q_n^{\ell_n}, R, q) \in \delta \text{ such that } t \in q
\end{equation}

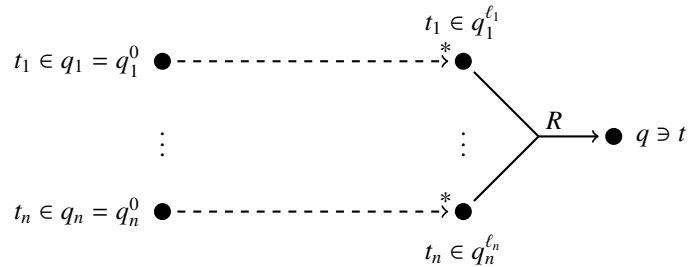
\begin{center}
\begin{tikzpicture}[inner sep=0, thick, outer sep=1mm, label distance=0mm]
\tikzset{nod/.style={draw,inner sep=.7mm,fill=black,circle}}
\draw (0,0) node(11)[nod,label={180:$t_1 \in q_1 = q_1^0$}]{};
\draw (4,0) node(1l)[nod,label={90:$t_1 \in q_1^{\ell_1}$}]{};
\draw (0,-2) node(n1)[nod,label={180:$t_n \in q_n = q_n^0$}]{};
\draw (4,-2) node(nl)[nod,label={-90:$t_n \in q_n^{\ell_n}$}]{};
\draw (5,-1) node(c)[outer sep=0,minimum size=0,label={45:$R$}]{};
\draw (6,-1) node(q)[nod,label={0:$q \ni t$}]{};
\draw (0,-1) node{$\vdots$};\draw (4,-1) node{$\vdots$};
\draw (11) edge[dashed,->] node[above,pos=.99]{$*$} (1l);
\draw (n1) edge[dashed,->] node[above,pos=.99]{$*$} (nl);
\draw (c) edge (1l) edge (nl);
\draw (c) edge[->] (q);
\end{tikzpicture}
\end{center}
\end{Definition}

\begin{Proposition}\label{prop:completeness}
For every complete proof tree automaton $\A$ on $\K$, $\D(\K) \subseteq \L(\A)$.
\end{Proposition}

\begin{proof}
Set $\A = (\K, Q, \delta, \delta\eps)$ a complete PTA. Let $D = (\pi, \lambda)$ be a derivation of $\K$. Let's build a run $\gamma : \pi \to Q$ by induction on $\pi$. We keep the invariant that if $\lambda(\varepsilon) = (t, R)$, then $\gamma(\varepsilon) = q$ (a word of length $1$), where $q$ is a conclusion state and $t \in q$.

A derivation is not empty.

\begin{itemize}
	\item Suppose $\pi = \set{\varepsilon}$ and $\lambda(\varepsilon) = (t, R)$. As $D$ is a derivation of $\K$, $(\varepsilon, t) \in R$. By completeness, there exists a state $q$ such that $(\varepsilon, R, q) \in \delta$ and $t \in q$. Therefore, by taking $\gamma(\varepsilon) = q$, $\gamma$ is a run of $\A$ on $D$ and $q$ is a conclusion state.
	
	\item Now suppose $D = (t, R) (D_1,..., D_n)$. Let's note $\lambda(j) = (t_j, R_j)$. By assumption, we have $(t_1...t_n, t) \in R$. By induction hypothesis, we can construct the runs $\gamma_j : \pi_j \to Q$ such that $\gamma_j(\varepsilon) = q_j^0$, where $q_j^0$ is a conclusion state and $t_j \in q_j^0$. As $\A$ is complete, there exists a state $q$ and $\varepsilon$-paths $q_j^0... q_j^{\ell_j} \in \varepsilon P_{\A}(t_j, q_j^0)$ such that $(q_1^{\ell_1}...q_n^{\ell_n}, R, q) \in \delta$ and $t \in q$. Thus, we can take the map $\gamma : \pi \to Q$ defined by
	
	\begin{align*}
	\gamma (\varepsilon) = & ~ q \\
	\gamma (j) = & ~ q_j^0... q_j^{\ell_j} & \text{for all $1 \leq j \leq n$} \\
	\gamma (j\nu) = & ~ \gamma_j(\nu) & \text{if $\nu$ is a node of $\pi_j$ except its root}
	\end{align*}
	
	It is straightforward to check that $\gamma$ is an execution on $\A$ and $q$ is a conclusion state.
\end{itemize}
As all executions are accepting in a PTA, we have $D \in \L(\A)$.
\end{proof}

%

\begin{Example}
	Figures \ref{fig:arthm-3} and \ref{fig:arthm-4} depict PTA of $\K_{\Arthm}$ which are not complete because there lacks a relation $\odd \xrightarrow{\Incr} \even$ and $\leone \darrow \even$ (or $ \leone \darrow 0$) respectively.
	
	We have $\L(\A_{\Arthm}^3) = \emptyset$. However, we still have the language inclusion for $\A_{\Arthm}^4$: $\K$, $\D(\K_{\Arthm}) \subseteq \L(\A_{\Arthm}^4)$, because we can always start a run with $ \xrightarrow{0} 0$ instead of $\xrightarrow{0} \leone$.
	
	Therefore, completeness is not a necessary condition to obtain this language inclusion.
\end{Example}

\begin{figure}[ht]
\begin{minipage}{.5\textwidth}
	\centering
\begin{tikzpicture}[thick,>=latex,outer sep=1mm]
\draw[circle] (0.5,-1) node[draw](0){$\zero$};
\draw[circle] (2,0) node[draw](e){$\even$};
\draw[circle] (4,0) node[draw](o){$\odd$};
\draw (1.5,.5) node[inner sep=0,outer sep=0,minimum size=0](i1){};
\draw (4.5,-.5) node[inner sep=0,outer sep=0,minimum size=0](i2){};
\draw (4.5,.8) node[inner sep=0,outer sep=0,minimum size=0](i3){};
\draw (-1,-1) node(i){};
\draw[->] (i) edge node[above]{$0$} (0) ;
\draw[dashed,->] (0) edge (e) ;
\draw[->] (e) edge[bend left] node[above]{$\Incr$} (o) ;
\draw (e) edge[out=145] (i1) edge[out=110] (i1)
edge[out=90,bend left=80] (i3) ;
\draw (o) edge[out=90] (i3)
edge[out=-40,bend left] (i2) edge[bend right,out=-40] (i2) ;
\draw[->] (i1) edge[bend right=100,in=-140] node[above left]{$\Add$} (e)
(i2) edge[bend left=70] node[above]{$\Add$} (e)
(i3) edge[bend left] node[right]{$\Add$} (o) ;
\end{tikzpicture}
	\captionof{figure}{Non complete PTA $\A^3_{\Arthm}$ of $\K_{\Arthm}$}
	\label{fig:arthm-3}
\end{minipage}
\hfill
\begin{minipage}{.45\textwidth}
	\centering
\begin{tikzpicture}[thick,>=latex,outer sep=1mm]
\draw[circle] (0.5,-1) node[draw](0){$\zero$};
\draw[circle] (-0.5,0.5) node[draw](l){$\leone$};
\draw[circle] (2,0) node[draw](e){$\even$};
\draw[circle] (4,0) node[draw](o){$\odd$};
\draw (1.5,.5) node[inner sep=0,outer sep=0,minimum size=0](i1){};
\draw (4.5,-.5) node[inner sep=0,outer sep=0,minimum size=0](i2){};
\draw (4.5,.8) node[inner sep=0,outer sep=0,minimum size=0](i3){};
\draw (-1,-1) node(i){};
\draw (-2,0.5) node(j){};
\draw[->] (i) edge node[above]{$0$} (0) ;
\draw[->] (j) edge node[above]{$0$} (l) ;
\draw[dashed,->] (0) edge (e) ;
\draw[->] (e) edge[bend left] node[above]{$\Incr$} (o) ;
\draw (e) edge[out=145] (i1) edge[out=110] (i1)
edge[out=90,bend left=80] (i3) ;
\draw[->] (o) edge[bend left] node[above]{$\Incr$} (e) ;
\draw (o) edge[out=90] (i3)
edge[out=-40,bend left] (i2) edge[bend right,out=-40] (i2) ;
\draw[->] (i1) edge[bend right=100,in=-140] node[above left]{$\Add$} (e)
(i2) edge[bend left=70] node[above]{$\Add$} (e)
(i3) edge[bend left] node[right]{$\Add$} (o) ;
\end{tikzpicture}
	\captionof{figure}{Non complete PTA $\A^4_{\Arthm}$ of $\K_{\Arthm}$, with $\leone = \set{0,1}$}
	\label{fig:arthm-4}
\end{minipage}
\end{figure}


\subsection{Totality and canonical PTA}\label{sec:other-properties}

Given any calculus $\K = (\Sign, \Varia, \R)$, the trivial consistent and complete PTA of $\K$ is given by only one state $q$ and the transitions $(\underbrace{q...q}_{n}, R_n, q)$ for every rule $R_n \in \R$ of arity $n$. This is clearly not what we want. It seems better to have finer source nodes (w.r.t. set inclusion) so we can better identify the hypotheses of rules on the graphical representation.

The goal of this section is to define a canonical PTA for every calculus, which have good properties.

We express the intuition of source fineness by totality of transitions. A transition labelled by $R$ is total if the cartesian product of its sources is included in the definition domain of $R$.

\begin{Definition}\label{def:totality}
In a PTA $\A = (\K, Q, \delta, \delta\eps)$, a transition $(q_1,...,q_n, R, q) \in \delta$ is total if

\begin{equation}
\text{for every } t_1 \in q_1,..., t_n \in q_n \text{, there exists } t \in q \text{ such that } (t_1...t_n, t) \in R
\end{equation}
\end{Definition}

\begin{Definition}
	A PTA is total if all its rule transitions are total.
\end{Definition}

Totality restrict the sources of rule transitions to be not too large. This property of PTA is directly related to a property of 

\begin{Definition}\label{def:modular}
A rule $R$ of arity $n$ is called \textbf{modular} if $\dom R  = \dom[1] R \times ... \times \dom[n] R$.
\end{Definition}

\begin{Remark}
Modularity gives interesting properties to the derivations of a term deduction system. When rules are presented as schematic rules (see appendix \ref{sec:schematic}, modularity is obtained as soon as there is no meta-variable common to different hypotheses. This does not hold in most traditional natural deduction or sequent calculi (e.g. introduction of $\wedge$). However, this holds for cut-free display calculi (see \cite{Greco_etal_LL:2016} for linear logic), as structural rules are always unary.
\end{Remark}

The canonical PTA of a calculus $\K$ is a simple way to crate a PTA of $\K$ with interesting properties.

\begin{Definition}\label{def:canonical-pta}
Given a calculus $\K = (\Sign, \Varia, \R)$, we defined the canonical PTA of $\K$ to be the PTA $\A = (\K, Q, \delta, \delta\eps)$ with
\begin{equation}
\begin{array}{rl}
     p_i^R = & \dom[i] R \\
     p_0^R = & \codom R \\
     Q = & \cset{p_i^R}{\exists n, \exists i \in \interv{0,n}, \exists R \in \R_n} \\
     \delta = & \cset{(p_1^R...p_n^R, R, p_0^R)}{\exists n, \exists R \in \R_n} \\
     \delta\eps = & \cset{(p_0^R, p_i^{R'})}{\exists n, \exists i \in \interv{1,n}, \exists R \in \R, R' \in \R_n.~ p_0^R \cap p_i^{R'} \neq \emptyset}
\end{array}
\end{equation}
\end{Definition}

\begin{Proposition}
The canonical PTA $\A$ of $\K$ is consistent and complete. Moreover, if every rule of $\K$ is modular, then $\A$ is total.
\end{Proposition}

\begin{proof}
	Using the same notation.
\begin{itemize}
	\item Given any term $t \in \codom R$, and rule $(q_1...q_n, R, q)$, we have $t \in q$ by construction, so $\A$ is consistent.
	
	\item If $R \in \R$ and $(t_1...t_n, t) \in R$, then for every conclusion states $q_1...q_n$, we have the $\varepsilon$-paths $q_ip_i^R$: $(q_i, p_i^R) \in \delta\eps$ and $(p_1^R...p_n^R, R, p_0^R) \in \delta$ by construction, and $t \in p_0^R$ also, hence $\A$ is complete.
	
	\item Suppose $R \in \R$ is modular. If $(q_1...q_n, R, q) \in \delta$ and $t_i \in q_i$ for all $i$, then by modularity, there exists $t \in \codom R$ such that $(t_1...t_n, t) \in R$. As $t \in q$ by construction, the transition $(q_1...q_n, R, q)$ is total. Therefore, if every rule of $\K$ is modular, $\A$ is total.
\end{itemize}
\end{proof}

\begin{Example}\label{ex:ptg-impl}
	This PTA is close to but not equal to the canonical PTA $\A_{\Can}$ for $\ImpL$. For example, in $\A_{\Can}$, there is an $\epsilon$-edge from $\Delta, \phi \vdash \psi$ to $\Delta \vdash \phi \to \psi$ because their set have a non-empty intersection, the first is the target of a rule edge and the second is the source of a rule edge. However, there exists an $\epsilon$-path from $\Delta, \phi \vdash \psi$ to $\Delta \vdash \phi \to \psi$ (the one passing by $\Gamma \vdash \phi$), and it preserves the terms, i.e. 
	\[\inst{\Delta, \phi \vdash \psi} ~\cap~ \inst{\Delta \vdash \phi \to \psi} ~~=~~ \inst{\Delta, \phi \vdash \psi} ~\cap~ \inst{\Gamma \vdash \phi} ~\cap~  \inst{\Delta \vdash \phi \to \psi}\]
	
	This correspondence holds for the other cases of canonical $\epsilon$-edges. In this sense, we can say that $\A_{\ImpL}$ is equivalent to $\A_{\Can}$.
\end{Example}

\section{Comparison to other systems}\label{sec:comparison}

In this section, we qualitatively compare proof tree automata with other graphical language classes related to logic: string diagrams and proof nets.

\subsection{String diagrams}

String diagrams \cite{Selinger:2011} are graphical languages aiming at representing equations in monoidal categories.

Informally, a category is a class of objects and morphisms connecting these objects. A monoidal category is a category having a tensor product, i.e. an operation giving an object $A \otimes B$ for any object $A$ and $B$, and which can also operate on morphisms. This intuitively corresponds to a composition law.

Let $\C$ be a class of monoidal categories for which there exists such a graphical language $\L$. This language explains how to construct a string diagram and how string diagrams can be rewritten into others, thus expressing equivalence between them.

The gist of a string diagram is as follows: objects are strings, tensor product of strings is juxtaposition of these strings, and morphisms are transformation of strings, represented by nodes having a domain (inputs) and a codomain (outputs).

For several logics (e.g. linear logic and some of its fragments), a class of monoidal categories exists w.r.t. which the calculus is sound and complete. In such a case, we can express a logical formula as a an object, and a judgment as a morphism. Thus, a judgment is representable by a string diagram. The rules of the calculus can be translated as equalities between string diagrams. Thus, a derivation corresponds to the rewriting of simple strings to a more complex geometrical arrangement of strings and nodes.

As they are graphical, string diagrams may represent certain properties as intuitive as they fit with our common perception of space. This more tangible representation allows us to better identify the key features of some derivation while ignoring structural ``uninteresting'' properties, like associativity.

On of these key features is information flow: how an object evolves during a derivation. It is easy to follow the string an quickly see that two formula occurrences actually correspond to the same string.

In classical logic, conjunction is a tensor product (it is actually a Cartesian product), and the other connectives can be represented by this product or some other typical objects (e.g. a coproduct for disjunction). But this good correspondence does not always easily extend when we add or remove specific properties of these connectives.

\subsection{Proof nets}

A proof structure \cite{Girard:1987} of type $t$ is a graph where vertices are signed formula occurrences of the sequent $t$ and edges either connect formula occurrences to their direct sub-formulas, or two occurrences of opposite sign of the same formula (axiom links). Proof nets are correct proof structures, i.e. proof structures with valid axiom links. In particular, there exists a proof net of type a given end sequent iff this sequent is derivable.

Rules can be translated as simple operations on proof nets. Given a derivation of end sequent $t$, we can thus construct a proof net of type $t$. But the goal of proof nets is to avoid traditional proof search.

A correction criterion establishes \cite{Danos_Regnier:1989} that a proof structure can be proven to be a prof net if a certain graph property holds. This criterion is often computed by applying graph rewriting rules.


As proof structures rely on subformulas, proof nets can also display information flow and axiom links.

\subsection{Common differences}

There two points which these two classes have in common but differ from proof tree graphs (PTGs) :
\begin{enumerate}
	\item the kind of object represented
	\item the existence of a general creation recipe
\end{enumerate}

\subsubsection{The kind of object represented}

These three classes have in common that they represent an object which denotes one derivation or a set of derivations of a given end sequent. 

On the contrary, PTGs show a broader map of a whole calculus. Derivations of any end sequent can be illustrated by their run in the automaton.

The focus of PTGs is rather on the relationships between the sequents, e.g. accessibility relations. Therefore, it is unfortunately harder to spot particular properties of single derivations, in particular axiom link and information flow.

To this extend, PTGs appear as a complementary tool to string diagrams and proof nets, rather than a competitor.

\subsubsection{General creation recipe}

Another common point of string diagrams and proof nets is their ungeneralizable dependence on the calculus. There may exist a intuitive, human-comprehensible description of what string diagrams (resp. proof nets) are in general, independently from any calculus. However, there is no general recipe allowing us to create a string diagram (resp. proof net) language $\L_L$ given \textit{any} logic $L$. The language $\L_L$ has to be specifically created knowing the peculiarities of $L$, and this construction may not always extend to other (even closely related) logics.

On the contrary, PTGs can represent any logic and even every term deduction system. Moreover, as we showed, there is always a canonical PTG with interesting properties.

However, PTGs are currently limited to the formalism of this paper: calculi having no term equivalence (in particular, no structure like multisets, so no hypersequent), no binding connective, no side conditions and only sort-consistent rules.


We have to point out that this argument about general recipe might get less valid in the future, as new ways of designing string diagrams or proof nets might arise.


\section{PTA as monoidal refinement systems}

In PTA, the control relations $\nabla$ and $\nabla\eps$, crucial to restrict runs on terms, may appear ad-hoc from an automata-theoretic viewpoint. In this part we argue that they naturally appear if we see PTA as monoidal refinement systems.

A PTA $\A$ can be decomposed into a calculus part and a graph part\footnote{Note that, considering the PTG as a traditional tree automaton, this decomposition emphasizes how PTA, and more generally \NCTAe, are related to tree automata: by parameterizing them by a monoidal category of derivations.}, viewed as monoidal categories $D_{\A}$ and $\T$ respectively. Morphisms of $\D_{\A}$ are runs on derivations, and morphisms of $\T$ are hyperwalks. $\A$ gives rise to a forgetful functor $U : \D_{\A} \to \T$, erasing terms to only keep states. This is illustrated in Fig.~\ref{fig:u}.

Control relations encode the conditions under which a hyperwalk is a run in $\A$. Therefore, the refinement system $U$ allows us to rephrase this by asking for an antecedent of such a hyperwalk. This constitutes theorem \ref{thm:antecedent}

\subsection{Categorical setting}

A refinement system \cite{Mellies_Zeilberger:2015} is a functor $U : \D \to \T$ between two categories. Intuitively, the object of $\T$ are types and the objects of $\D$ are called refinement types. If $U(S) = A$, we say that $S$ refines $A$. The morphisms of $\D$ are derivations and the morphisms of $\T$ are often computational terms (e.g. $\lambda$-terms). If $S \xrightarrow{\alpha} T$ is a morphisms of $\D$, then $U$ maps $\alpha$ to $U(S) \xrightarrow{f} U(T)$ with $f = U(\alpha)$. We write this as the typing judgment $S \xRightarrow[f]{\alpha} T$.


We say that a refinement system is monoidal if $\D$, $\T$ and $U$ are monoidal, i.e. $U$ respects the tensor product. We write $I$ the identity element.

First, let's define the categories $\D$, $\T$ and the functor $U$. Set $\K = (\Sign, \Varia, \R)$ a calculus and $\A = (\K, Q, \delta, \delta\eps)$ a PTA on $\K$.

\begin{Definition}
We define $\D$ as the free monoidal category generated by $\cset{(t, T) \in \T(\Sign, \Varia) \times \wp(\T(\Sign, \Varia))}{t \in T}$ as objects, and by the arrows
\begin{equation}
\begin{array}{rrcll}
& (t_1, T_1) \otimes ... \otimes (t_n, T_n) & \xrightarrow{R} & (t, T) & \text{ such that } (t_1...t_n, t) \in R, \text{ for every rule } R \in \R \\
\text{and } & (t, T) & \xrightarrow{e_{T,T'}} & (t', T') & \text{ such that } t = t'
\end{array}
\end{equation}
as morphisms.
\end{Definition}

In $\D$, an object is a tensor product of terms (e.g. sequents of formulas) as part of a set of terms $T$. Morphisms are partial derivations. A morphism $I \xrightarrow{D} (t, T)$ means that $D$ is a derivation of $\K$ with root term $t$.

\begin{Definition}
We define $\T$ the free monoidal category generated by $Q$ as objects, and by
\begin{equation}
\begin{array}{rrcll}
& q_1 \otimes ... \otimes q_n & \xrightarrow{d} & q & \text{ for } d \in \delta \\
\text{and } & q & \xrightarrow{d} & q' & \text{ for } d \in \delta\eps
\end{array}
\end{equation}
as morphisms.
\end{Definition}

The category $\T$ embodies the automaton structure of $\A$, for morphisms are based on state transitions. A morphism of $\T$ is a hyperwalk on the underlying PTG of $\A$: it does not take instance terms and control relations into account.

Now we can define $U$ as a ``partial'' functor from $\D$ to $\T$.

\begin{Definition}
Set $\D_{\A}$ the subcategory of $\D$ restricted to tensor products of $\T(\Sign, \Varia) \times Q$ and morphisms
\begin{equation}
\begin{array}{rrcll}
& (t_1, T_1) \otimes ... \otimes (t_n, T_n) & \xrightarrow{R} & (t, T) & \text{ such that } (T_1...T_n, R, T) \in \delta \\
\text{and } & (t, T) & \xrightarrow{e_{T,T'}} & (t, T') & \text{ such that } (T, T') \in \delta\eps
\end{array}
\end{equation}

We define $U : \D_{\A} \to \T$ as the monoidal functor generated by taking
\begin{equation}
\begin{array}{rrcl}
U : & (t, q) & \mapsto & q \in Q \\
 & (t_1, q_1) \otimes ... \otimes (t_n, q_n) \xrightarrow{R} (t, q) & \mapsto & q_1 \otimes ... \otimes q_n \xrightarrow{(q_1...q_n, R, q)} q \\
 & (t, q) \xrightarrow{e_{q,q'}} (t, q') & \mapsto & q \xrightarrow{(q, q')} q'
\end{array}
\end{equation}
\end{Definition}

\begin{Proposition}
$U : \D_{\A} \to \T$ is a monoidal refinement system, by construction.
\end{Proposition}

$U$ is a forgetful functor. It erases terms and control.

%
%
%
%
%
%
%
%

Encoding $\A$ as a refinement system $U$ shines a light on the decomposition of a PTA as a partial map from a calculus $\D$ to a graph $\T$, as illustrated in Fig.~\ref{fig:u}.

\begin{figure}[ht]
	\centering
\begin{minipage}{.5\textwidth}
	\begin{tikzpicture}[thick,rectangle]
	\draw (0,0) node(c){Calculus} ;
	\draw (0,-3) node(g){Graph} ;
	\draw (3,-1) node(r){\tab{Derivation \& run\\$I \xrightarrow{D, \gamma} (t, q)$}} ;
	\draw (3,-3) node(h){\tab{Hyperwalk\\$I \xrightarrow{H} q$}} ;
	\draw (6,0) node(d){$\D$} ;
	\draw (6,-1) node(da){$\D_{\A}$} ;
	\draw (6,-3) node(t){$\T$} ;
	\draw[->] (c) edge node[left]{PTA} (g)
	(da) edge node[right]{$U$} (t) ;
	\draw[|->] (r) edge (h) ;
	\draw[right hook->] (da) edge (d) ;
	\end{tikzpicture}
	\captionof{figure}{Decomposition of a PTA $\A$ as a refinement system $U$ between a calculus and a graph.}
	\label{fig:u}
\end{minipage}
\hfill
\begin{minipage}{.45\textwidth}
	\hfill
\begin{tikzpicture}[thick]
\draw (0,0) node(i){} ;
\draw (0,-1) node[draw,ellipse](e){$\Add$} ;
\draw (-1,-2) node[draw,ellipse](1){$\Incr$} ;
\draw (-1,-3) node[draw,ellipse](11){$\varepsilon$} ;
\draw (-1,-4) node[draw,ellipse](111){$0$} ;
\draw (1,-2) node[draw,ellipse](2){$\Add$} ;
\draw (0.5,-3) node[draw,ellipse](21){$\Incr$} ;
\draw (0.5,-4) node[draw,ellipse](211){$\Incr$} ;
\draw (0.5,-5) node[draw,ellipse](2111){$\varepsilon$} ;
\draw (0.5,-6) node[draw,ellipse](21111){$0$} ;
\draw (1.5,-3) node[draw,ellipse](22){$\varepsilon$} ;
\draw (1.5,-4) node[draw,ellipse](221){$0$} ;
\draw (i) edge node[right]{$\odd$} (e)
	(e) edge node[above left]{$\odd$} (1)
		edge node[above right]{$\even$} (2)
	(1) edge node[left]{$\even$} (11)
	(11) edge node[left]{$\zero$} (111)
	(2) edge node[left]{$\even$} (21)
		edge node[right]{$\even$} (22)
	(21) edge node[left]{$\odd$} (211)
	(211) edge node[left]{$\even$} (2111)
	(2111) edge node[left]{$\zero$} (21111)
	(22) edge node[right]{$\zero$} (221)
;
\draw (-1,-5.5) node{$\T$} ; \draw (2,-5.5) node{$\T$} ;
\end{tikzpicture}
\captionof{figure}{String diagram representing a morphism $I \xrightarrow{H} \odd$ of hyperwalk $H$ on $\G_{\Arthm}$. $H$ is correct w.r.t. to a run on the derivation in Fig.~\ref{fig:tree-arthm}.}
\label{fig:string}
\end{minipage}
\end{figure}

\subsection{Correction of hyperwalk as validity of typing judgment}

In a refinement system, a typing judgment $S \xRightarrow[f]{} T$ is said to be valid if $f$ admits an antecedent. In our case, $f$ is a (generalized) hyperwalk on the underlying graph of $\A$. Validity amounts to the possibility to label that hyperwalk with terms form a derivation and which respect the restrictions the rules put on terms, i.e. the control relations.

\begin{Theorem}\label{thm:antecedent}
A typing judgment $I \xRightarrow[H]{} (t, q)$ is valid iff the hyperwalk $H$ is correct.
\end{Theorem}

\begin{proof}
	First, let us explicit how a morphism $I \xrightarrow{H} q$ of $\T$ can be seen as a hyperwalk and a morphism $I \xRightarrow{\gamma} (t, q)$ of $\D_{\A}$ as a run in $\A$.
	
	By tensor-atomic object we mean an object $A$ (resp. morphism $\alpha$) which is not of the form $A_1 \otimes A_2$ ($A_i \neq I$) (resp. $\alpha_1 \otimes \alpha_2$, $\alpha_i \neq \id_I$). By construction, atomic objects are states $q$ for $\T$ and pair $(t, T)$ for $\D_{\A}$, and atomic morphisms are elements of $\delta$ and $\delta\eps$ for $\T$ and rules $R$ or arrows $e_{T, T'}$ for $\D_{\A}$.
	
	By composition-atomic morphism $\alpha$ we mean that $\alpha$ is not of the form $\alpha_1 \circ \alpha_2$ with $\alpha_i \neq \id$.
	
	Given that $\D_{\A}$ and $\T$ are free categories, we can decompose a morphisms $A \xrightarrow{\alpha} B$ into $A = A_0 \xrightarrow{\alpha_1} ... \xrightarrow{\alpha_n} A_n = B$ with $\alpha_i$ composition-atomic. So we can proceed by induction on $n$.
	
	Moreover, as the target of all tensor-atomic morphisms is a tensor-atomic object, we can decompose composition-atomic morphisms $A \xrightarrow{\alpha} B_1 \otimes ... \otimes B_n$ in tensor-atomic morphisms $\alpha = \alpha_1 \otimes ... \otimes \alpha_n$, with $A_1 \xrightarrow{\alpha_1} B_i$ and $A = A_1 \otimes ... \otimes A_n$.
	
	Thus, any morphism $A \xrightarrow{\alpha} B$ can be seen as a forest of roots given by $B$. In particular, if $B$ is tensor-atomic and $A = I$, this forest is a tree with axioms at leaves. So we can extract from $\alpha$ a tree $\pi$. We write $A_{\nu}$ the tensor-atomic object at node $\nu$ (e.g. $A_{\varepsilon} = B$) and $\alpha_{\nu}$ the tensor-atomic and composition-atomic arrow of target $A_{\nu}$.
	
	In both $\T$ and $\D_{\A}$, we can squeeze $\varepsilon$-transitions by taking a subtree $\pi' \subseteq \pi$ such that $\nu \not\in \pi'$ iff $\alpha_{\nu}$ is an $\varepsilon$-transition (i.e. some $e_{T, T'}$ or $d \in \delta\eps$) or identity.
	\medskip
	
	In $\T$, we can view a morphism $I \to q$ as the term $H = (\pi', (\delta \times Q) (\delta\eps \times Q)^*, \lambda_H)$, where
	\begin{equation}
	\lambda_H(\nu) = (d, A_{\nu}) (d_1, A_{\nu_1}) ... (d_n, A_{\nu_m}) ~~~~\text{ if }~~~~ A_{\nu^1} \otimes ... \otimes A_{\nu^n} \xrightarrow{d} A_{\nu} \xrightarrow{d_2} A_{\nu_2} ... \xrightarrow{d_m} A_{\nu_m}
	\end{equation}
	with $\nu_m$ the longest prefix of $\nu$ such that $\nu_m = \varepsilon$ or the immediate prefix of $\nu_m$ belongs to $\pi'$.
	
	$H$ is a hyperwalk on the PTG of $\A$.
	\medskip
	
	In $\D_{\A}$, we can view a morphism $I \to (t, q)$ as the term $D = (\pi, (\T(\Sign, \Varia) \times \R, \lambda_D)$ together with a map $\gamma : \pi \to Q^+$, where
	\begin{equation}
	\begin{array}{rcll}
	\lambda_D(\nu) & = & (t_{\nu}, R) & \\
	\gamma(\nu) & = & q_{\nu} q_{\nu_1}...q_{\nu_m} & ~~\text{ if }~~~~ A_{\nu^1} \otimes ... \otimes A_{\nu^n} \xrightarrow{R} (t_{\nu}, q_{\nu}) \xrightarrow{e_2} (t_{\nu_2}, q_{\nu_2}) ... \xrightarrow{e_m} (t_{\nu_m}, q_{\nu_m})
	\end{array}
	\end{equation}
	with $\nu_m$ like above.
	
	$D$ is a derivation on $\K$ and $\gamma$ a run on $D$ in $\A$.
	\medskip
	
	Now, let us prove that, for $t \in q$, a morphism $I \xrightarrow{H} q$ admits an antecedent $I \xrightarrow{D, \gamma} (t, q)$ iff $H$ is correct. Proving this reformulation of the theorem would thus prove the theorem.
	
	$\boxed{\Rightarrow}$ Suppose $H$ admits an antecedent $I \xrightarrow{D, \gamma} (t, q)$. As $U$ maps tensor-atomic and composition-atomic morphisms to tensor-atomic and composition-atomic morphisms, $H, D$ and $\gamma$ are based on the same tree. Set $\nu \in \pi'$ of daughter $\nu^1,..., \nu^n \in \pi$ and $\lambda_D(\nu) = (t, R)$. Let us write
	\[ (t_{\nu^1}, q_{\nu^1}) \otimes ... \otimes (t_{\nu^n}, q_{\nu^n}) \xrightarrow{R'} (t_{\nu}, q_{\nu}) \xrightarrow{e_2} (t_{\nu_2}, q_{\nu_2}) ... \xrightarrow{e_m} (t_{\nu_m}, q_{\nu_m})\]
	
	the situation at $\nu$ in $\D_{\A}$. by definitions of $D$, we must have $t = t_{\nu}$ and $R = R'$. By definition of $H$ and $U$, we have
	\begin{align*}
	\lambda_H(\nu) & = (U(R), U((t_{\nu}, q_{\nu}))) (U(e_1), U((t_{\nu_2}, q_{\nu_2}))) ... (U(e_m), U((t_{\nu_m}, q_{\nu_m}))) \\
	& = ((q_{\nu^1}...q_{\nu^n}, R, q_{\nu}), q_{\nu}) ((q_{\nu}, q_{\nu_2}), q_{\nu_2}) ... ((q_{\nu_{m-1}}, q_{\nu_m}), q_{\nu_m}) \\
	\end{align*}
	
	By definition of $\gamma$, for all $j \leq m$, $\gamma(\nu^j)_{m_j} = q_{\nu^j}$ (with $m_j = \size{\gamma(\nu^j)}$), $\gamma(\nu)_1 = q_{\nu}$ and for all $1 \leq i \leq n$, $\gamma(\nu)_i = q_{\nu_i}$. Therefore we get
	\begin{align*}
	\lambda_H(\nu)
	& = ((\gamma(\nu^1)_{m_1}...\gamma(\nu^n)_{m_n}, R, \gamma(\nu)_1), \gamma(\nu)_1) ((\gamma(\nu)_1, \gamma(\nu)_2), \gamma(\nu)_2) ... ((\gamma(\nu)_ {m-1}, \gamma(\nu)_m), \gamma(\nu)_m) \\
	\end{align*}
	
	 as \eqref{eq:correction} of definition \ref{def:correction}. This shows that $H$ is correct w.r.t. $\A$ via $D$ and $\gamma$.
	\medskip
	
	 \rotatebox[origin=c]{180}{$\boxed{\Rightarrow}$} Conversely, suppose $H$ is correct w.r.t. $\A$. There exists a derivation $D$ and a run $\gamma$ on $D$ in $\A$ verifying equation \ref{eq:correction}. We can create a morphism $I \xrightarrow{\alpha} (t, q)$ from $D$ and $\gamma$ by reverting the process described above. Showing that $U(\alpha) = H$ is follows the same reasoning as the direct implication.
	 
	 \medskip
	 The equivalence is proven, which proves the theorem.
\end{proof}

Finally, let us mention an interesting point related to section \ref{sec:comparison}. Modelling PTA through monoidal refinement systems enables to use string diagrams to represent morphisms. As a consequence, string diagrams appear as possible unfoldings of PTA. An example hyperwalk as string diagram for PTA $\A_{\Arthm}$ is given in Fig.~\ref{fig:string}.

\section{Conclusion and open questions}

\subsection{Summary}

We presented a new class of tree automata called non-deterministic controlling tree automata with $\varepsilon$-transitions (\NCTAe). Compared to a non-deterministic tree automaton with $\varepsilon$-transitions, a \NCTAe $\A$ can run on trees which labels comprise an element $t$ of an infinite set called instance. There are two controlling relations, $\nabla$ for transitions and $\nabla\eps$ for $\varepsilon$-transitions, which set restrictions on instances involved in a transition of $\A$.

We define a proof tree automaton (PTA) $\A$ on a calculus (or any term deduction system) $\K$ to be a \NCTAe dependent on $\K$. We expose a sufficient completeness criteria to ensure that the language of $\A$ is equal to the derivation language of $\K$. This shows that proof tree automata are a tool able to reason about the derivations produced by a calculus.

A proof tree graph (PTG) is defined as the graphical representation of a proof tree automaton. This hypergraph brings a visual intuition about the ways the rules of a calculus are connected to each other by depicting rules as labelled hyperedges and sets of terms as vertices.

We compared this graphical language to string diagrams and proof nets. The main difference with these two concepts lies in the kind of object represented. String diagrams and proof nets represent a set of equivalent derivations of only one end sequent, whereas a proof tree graph represent the whole calculus.

Thus, proof tree automata and proof tree graphs appear as novel formal system and graphical language, shedding a new light on term deduction systems as finite state machines and graphs respectively.

\subsection{Open questions}

This contribution aims at setting a clear basis for the study of PTA and PTGs. Therefore, it leaves more open questions than answered ones.

\begin{enumerate}

\item An important aspect of finite state machines which is not mentioned at all here is complexity. Supposing the calculus is presented as a schematic calculus, the controlling relations just amount to unification between schematic terms. Efficient and linear algorithms \cite{Martelli_Montanari_unification:1982,Paterson_Wagman_unification:1976} solving unification exist. Thus, testing whether a tree is a derivation of this calculus by running a complete PTA on it can be performed in polynomial time under some reasonable hypotheses (e.g. asking that the backward proof-search does not increase the size of the terms too much). Working on general \NCTAe, we could imagine that the instances are equipped with a size function $\size{\cdot} : \Phi \to \mathbb{R}_+$ and that the controlling relations are polynomial w.r.t. the size of the instances, in order to analyse their complexity.

\item It may also be useful to come up with criteria which are necessary condition to the inclusion of languages as stated in propositions \ref{prop:soundness} and \ref{prop:completeness} respectively. To do so, we would probably need to design a good notion of accessibility of a rule in a calculus and of a state in a PTA.

\item From a combinatorial viewpoint, we can wonder whether PTA could help us compute the number of derivations of a given end sequent. As far as enumeration is concerned, PTA merit to be compared to inhabitation machines \cite[p. 33]{Barendregt_etal:2013}.

\item There may be different consistent and complete PTA on a given calculus. An interesting question is: How do they relate to each other? This question first raises an investigation on equivalence of PTA. Two PTA can be defined equivalent if they share the same language. But we could also imagine other (e.g. finer) relevant notions of equivalence, focussing on certain internal aspects of proof tree constructions. The question also raises an investigation on ordering PTA. We would like to be able to express that a PTA is finer than another one if we can map the states of the second ones to the first one while preserving the transitions. There is here something non-trivial to adapt from tree automata to define PTA morphisms, and more generally the category of PTA. Moreover, developing graph manipulation techniques as an operational way to compute equivalent PTG having a desirable property is certainly a research axis of practical use (e.g. see the comment at the end of example \ref{ex:ptg-impl}).

\item Another point of curiosity arises when considering correction of hyperwalks. The current formulation involves the existence of a run on a derivations, thus requiring to find some term labelling. However, could it be possible to express correction without ever mentioning terms, but rather by looking at the previous nodes of the hyperwalk? As subsidiary question: given a $\varepsilon$-edge $e$, could we design a criterion $P_e$ on hyperwalks, such that $P_e(H)$ iff $e$ can be used starting from the root of the hyperwalk $H$?

\end{enumerate}

Finally, given that in the case where all rules are unary, a term deduction system is just a term rewriting system, PTA and PTG are also usable a for term rewriting systems.


\printbibliography

\appendix

\section{Schematic PTA}\label{sec:schematic}

In this section, we give a closer look at the implementation of proof tree graphs as finite representations. As most calculi in the literature are given as schematic calculi, it sounds reasonable to define a schematic PTA as a finite machine using rule names and schematic terms instead of infinite sets.

\subsection{Schematic calculi}

Schematic terms and schematic rules\footnote{The words \textit{schematic} and \textit{instance} are taken from \cite{Greco_etal_LE:2018}.} are a common ways to represent a calculus with a finite number of symbols.

Set $\Sign$ a signature and $\Varia$ a variable set on the same sort set $\S$. We set $\Meta = (\S, \M, \trg)$ a variable set with a countably infinite number of elements (called meta-variables) of each sort such that $\M \cap \V = \emptyset$.

The set of closing meta-variable substitutions $\sigma : \Meta \nrightarrow \T(\Sign, \Varia)$ is written $\metasub$.

\begin{Definition}
	A \textbf{schematic term} is a term $\term \in \T(\Sign, \Varia, \Meta)$ made of connectives, variables and meta-variables.
	
	A \textbf{schematic rule} is a pair $\rulsch = (\term_1... \term_n, \term) \in \T(\Sign, \Varia, \Meta)^* \times \T(\Sign, \Varia, \Meta)$.
	
	The instance set of a schematic term $\term$ is the set of terms $ \inst{\term} = \cset{\term \sigma}{\sigma \in \metasub}$.
	
	An instance of $\rulsch$ by a meta-variable substitution $\sigma \in \metasub$ is the instance rule $\rulsch \sigma = ((\term_1 \sigma)... (\term_n \sigma), \term \sigma)$. The rule associated with $\rulsch$ is $\inst{\rulsch}$, the set of instances of $\rulsch$ by $\sigma$, for all $\sigma \in \metasub$.
\end{Definition}

The problem of checking whether a term (resp. a rule) is an instance of a schematic term (resp. rule), i.e. whether it belongs to its instance set, can be decided by unification \cite{Herbrand:1930}.

\begin{Definition}\label{def:sch-calc}
	A \textbf{schematic calculus} is a quintuple $\Calc = (\Sign, \Varia, \Meta, \Names, \Rules)$ where
	\begin{itemize}
		\item $\Sign$ a signature
		\item $\Varia$ a variable set on the same sort set $\S$
		\item $\Meta$ is a set of meta-variables on $\S$
		\item $\Names$ is a finite set of elements called rule names
		\item $\Rules : \Names \to \T(\Sign, \Varia, \Meta)^* \times \T(\Sign, \Varia, \Meta)$ maps every rule name to a schematic rule
	\end{itemize}
	
	The instance calculus associated to $\Calc$ is $\inst{\Calc} = (\Sign, \Varia, \inst{\Rules})$, with $\inst{\Rules} = \cset{\inst{\Rules(N)}}{N \in \Names}$
\end{Definition}

\begin{Definition}\label{def:sch-der}
	A \textbf{schematic derivation} in a schematic calculus $\Calc = (\Sign, \Varia, \Meta, \Names, \Rules)$ is a rooted labelled tree $\der = (\pi, {\T(\Sign, \Varia) \times \Names}, \lambda)$ such that
	\begin{itemize}
		\item for all $\nu \in \pi$ of daughters $\nu_1,..., \nu_n$, if $\lambda(\nu) = (t, N)$ and $\lambda(\nu_i)= (t_i, N_i)$ for all $i$, then $(t_1,...,t_n, t) \in \inst{\Rules(N)}$
	\end{itemize}

	The instance of a schematic derivation is the derivation $D = (\pi, \T(\Sign, \Varia) \times \inst{\Rules}, \lambda')$ where for all $\nu \in \pi$, if $\lambda(\nu) = (t, N)$ then $\lambda'(\nu) = (t, \inst{\Rules(N)})$.
	
	$\D(\Calc)$ is the set of schematic derivations of $\Calc$.
	If $\mathbf{L}$ is a set of schematic derivations, $\inst{\mathbf{L}}$ denotes $\cset{\inst{\der}}{\der \in \mathbf{L}}$.
\end{Definition}

\begin{Proposition}\label{prop:inst}
	If $\Calc$ is a schematic calculus, then $\inst{\D(\Calc)} = \D(\inst{\Calc})$.
\end{Proposition}

\begin{proof} We use the same notation as above.
\begin{itemize}
	\item Set $D \in \inst{\D(\Calc)}$. There exist $\der \in \D(\Calc)$ such that $\inst{\der} = D$. By definitions \ref{def:sch-calc} and \ref{def:sch-der}, it is clear that $\inst{\der}$ is a derivation of $\inst{\Calc}$, so $D \in \D(\inst{\Calc})$.
	
	\item Set $D = (\pi, \lambda) \in \D(\inst{\Calc})$. Let $\der = (\pi, \lambda')$ be a schematic derivation where $\lambda'$ is constructed as follows. Given a node $\nu$ with $\lambda(\nu) = (t, R)$, we set $\lambda'(\nu) = (t, N)$ with $N$ an element of $\Names$ such that $\Rules(\Names) = R$. $N$ exists because $R$ is a rule of $\inst{\Calc}$ and $\Names$ is finite. It is straightforward to check that $\der$ is a schematic derivation of $\Calc$ and $\inst{\der} = D$. Consequently $D \in \inst{\D(\Calc)}$.
\end{itemize}
\end{proof}

\subsection{Schematic PTA}

In practice, it is useful to draw a PTG with schematic terms as vertex labels and rules names as edge labels. Therefore, we introduce the simple notion of schematic PTA and PTG.

\begin{Definition}
	Given a schematic calculus $\Calc = (\Sign, \Varia, \Meta, \Names, \Rules)$, a \textbf{schematic PTA} on $\Calc$ is a tuple $\Aut = (\Calc, \Names, \States, \trans, \trans\eps)$ where
	\begin{itemize}
		\item $\States \subseteq_f \T(\Sign, \Varia, \Meta)$ \hfill (states are schematic terms term)
		\item $\trans\subseteq \bigcup_{n \in \nats} \States^n \times \Names_n \times \States$ \hfill (transitions are labelled by rule names)
		\item $\trans\eps \subseteq \States \times \States$
	\end{itemize}
	
	Here $\Names_n$ is the set of names mapped to schematic rules of arity $n$.
\end{Definition}

\begin{Definition}
	Using the same notation, the instance of $\Aut$ is $\inst{\Aut} = (A, \Phi, \inst{\States}, \inst{\trans}, \nabla, F, \inst{\trans\eps}, \nabla\eps)$ where
	\begin{equation}
	\begin{array}{rcl}
	A &=& (\S, \inst{\Rules}, \src^A, \trg^A) \\
	\Phi &=& \T(\Sign, \Varia) \\
	\inst{\States} &=& \cset{\inst{\term}}{\term \in \States} \\
	\inst{\trans} &=& \cset{(\inst{\term_1}...\inst{\term_n}, \inst{\Rules(N)}, \inst{\term})}{(\term_1...\term_n, N, \term) \in \trans}\\
	(t_1...t_n, R, t) \in \nabla & \text{ iff } & (t_1...t_n, t) \in R \\
	F &=& \inst{\States} \\
	\inst{\trans\eps} &=& \cset{(\inst{\term_1}, \inst{\term_2})}{(\term_1, \term_2) \in \trans\eps} \\
	(t, q) \in \nabla\eps & \text{ iff } & t \in q
	\end{array}
	\end{equation}
	Again, $A$ is a signature by definition \ref{def:rule}.
\end{Definition}

\begin{Proposition}
	The instance of a schematic PTA on $\Calc$ is a PTA on $\inst{\Calc}$.
\end{Proposition}

\begin{proof}
	It is straightforward to check that for every $q \in \inst{\States}$, $q \in \wp(\Phi)$. The resting requirements are clearly met.
\end{proof}

\begin{Definition}
	A schematic PTA is consistent (resp. complete, total) if its instance PTA is consistent (resp. complete, total) w.r.t. the instance of its calculus.
\end{Definition}

\begin{Definition}\label{def:sch-run}
	A \textbf{run of a schematic} PTA $\Aut = (\Calc, \Names, \States, \trans, \trans\eps)$ on a schematic derivation $\der = (\pi, \lambda)$ is a map $\run : \pi \to \States^+$ such that, for every node $\nu \in \pi$ labelled by $(t, N)$,
	\begin{itemize}
		\item if the daughters of $\nu$ are $\nu_1,..., \nu_n$ then $(\run(\nu_1)_{m_1}... \run(\nu_n)_{m_n}, N, \run(\nu)_0) \in \trans$, where $m_j = \size{\run(\nu_j)}$ \hfill (transition)
		
		\item by noting $ \lambda(\nu_j) = (t_j, a_j)$ for $1 \leq j \leq n$, then $(t_1... t_n, t) \in \inst{\Rules(N)}$ \hfill (control)
		
		\item for every $0 \leq i < \size{\run(\nu)} - 1$, $(\run(\nu)_i, \run(\nu)_{i+1}) \in \trans\eps$, \hfill ($\varepsilon$-transitions)
		
		\item and $t \in \inst{\run(\nu)_{i+1}}$ \hfill ($\varepsilon$-control)
	\end{itemize}

	We write $\L(\Aut)$ the set of derivations recognized by a run on $\Aut$.
\end{Definition}

\begin{Proposition}
	If $\Aut$ is a schematic PTA on $\Calc$, then $\inst{\L(\Aut)} = \L(\inst{\Aut})$.
\end{Proposition}

\begin{proof} We use the same notation.
\begin{itemize} 
	\item Set $D = (\pi, \lambda) \in \inst{\L(\Aut)}$. There exists $\der \in \L(\Aut)$ such that $\inst{\der} = D$, so there exists a run $\run : \pi \to \States$ on $\der$. We define $\gamma : \pi \to \inst{\States}$ to be, for every $\nu \in \pi$, $\gamma(\nu) = \inst{\run(\nu)}$. Checking that every item of definition \ref{def:sch-run} on $\run$ translates to a respective item of definition \ref{def:run-NCTAe} on $\gamma$, is left to the reader. It yields that $\gamma$ is a run of $\inst{\Aut}$, hence $D \in \L(\inst{\Aut})$.
	
	\item Set $D = (\pi, \lambda) \in \L(\inst{\Aut})$ and $\gamma : \pi \to \inst{\States}$ a run recognizing $D$. Similarly to the second part of the proof of proposition \ref{prop:inst}, we can construct a schematic derivation $\der$ such that $\inst{\der} = D$ and a run $\run : \pi \to \States$ on $\der$, by taking a name for each rule and a schematic term in $\States$ for each $q \in \inst{\States}$. Therefore, $D \in \inst{\L(\Aut)}$.
\end{itemize}
\end{proof}

\begin{Definition}
	The schematic proof tree graph of a schematic PTA $(\Calc, \Names, \States, \trans, \trans\eps)$ is the \DHGd $\mathbf{G} = (\States, \trans, \Names, \trans\eps)$.
\end{Definition}

Figure \ref{fig:pta-ImpL} actually shows the underlying schematic PTG of the PTA $\A_{\ImpL}$.

\end{document}